# Magnetic properties in ultra-thin 3d transition metal alloys II: Experimental verification of quantitative theories of damping and spin-pumping

Martin A. W. Schoen,[1,2]* Juriaan Lucassen,[3] Hans T. Nembach,[1] Bert Koopmans,[3] T. J. Silva,[1] Christian H. Back,[2] and Justin M. Shaw[1]

[1]Quantum Electromagnetics Division, National Institute of Standards and Technology, Boulder, CO 80305, USA
[2]Institute of Experimental and Applied Physics, University of Regensburg, 93053 Regensburg, Germany
[3]Department of Applied Physics, Eindhoven University of Technology, 5600 MB Eindhoven, The Netherlands

Dated: 01/05/2017

*Corresponding author: martin1.schoen@physik.uni-regensburg.de

## Abstract

A systematic experimental study of Gilbert damping is performed via ferromagnetic resonance for the disordered crystalline binary 3$d$ transition metal alloys Ni-Co, Ni-Fe and Co-Fe over the full range of alloy compositions. After accounting for inhomogeneous linewidth broadening, the damping shows clear evidence of both interfacial damping enhancement (by spin pumping) and radiative damping. We quantify these two extrinsic contributions and thereby determine the intrinsic damping. The comparison of the intrinsic damping to multiple theoretical calculations yields good qualitative and quantitative agreement in most cases. Furthermore, the values of the damping obtained in this study are in good agreement with a wide range of published experimental and theoretical values. Additionally, we find a compositional dependence of the spin mixing conductance.

# 1 Introduction

The magnetization dynamics in ferromagnetic films are phenomenologically well described by the Landau-Lifshitz-Gilbert formalism (LLG) where the damping is described by a phenomenological damping parameter $\alpha$.[4,5] Over the past four decades, there have been considerable efforts to derive the phenomenological damping parameter from first principles calculations and to do so in a quantitative manner. One of the early promising theories was that of Kamberský, who introduced the so-called breathing Fermi surface model[6–8]. The name "breathing Fermi surface" stems from the picture that the precessing magnetization, due to spin-orbit coupling, distorts the Fermi surface. Re-populating the Fermi surface is delayed by the scattering time, resulting in a phase lag between the precession and the Fermi surface distortion. This lag leads to a damping that is proportional to the scattering time. Although this approach describes the so-called conductivity-like behavior of the damping at low temperatures, it fails to describe the high temperature behavior of some materials. The high temperature or resistivity-like behavior is described by the so-called "bubbling Fermi surface" model. In the case of energetically shifted bands, thermal broadening can lead to a significant overlap of the spin-split bands in $3d$ ferromagnets. A precessing magnetization can induce electronic transitions between such overlapping bands, leading to spin-flip processes. This process scales with the amount of band overlap. Since such overlap is further increased with the band broadening that results from the finite temperature of the sample, this contribution is expected to increase as the temperature is increased. This model for interband transition mediated damping describes the resistivity-like behavior of the damping at higher temperatures (shorter scattering times). These two damping processes are combined in a torque correlation model by Gilmore, et al.[9], as well as Thonig, et al.[10], that describes both the low-temperature (intraband transitions) and high-temperature (interband transitions) behavior of the damping. Another approach via scattering theory was successfully implemented by Brataas, et al.[11] to describe damping in transition metals. Starikov, et al.,[2] applied the scattering matrix approach to calculate the damping of $Ni_xFe_{1-x}$ alloys and Liu, et al.,[12] expanded the formalism to include the influence of electron-phonon interactions.

A numerical realization of the torque correlation model was performed by Mankovsky, *et al.*, for $Ni_xCo_{1-x}$, $Ni_xFe_{1-x}$, $Co_xFe_{1-x}$, and $Fe_xV_{1-x}$[1]. More recently, Turek, et al.,[3] calculated the damping for $Ni_xFe_{1-x}$ and $Co_xFe_{1-x}$ alloys with the torque-correlation model, utilizing non-local torque correlators. It is important to stress that all of these approaches consider only the intrinsic damping. This complicates the quantitative comparison of calculated values for the damping to experimental data since there are many extrinsic contributions to the damping that result from sample structure, measurement geometry, and/or sample properties. While some extrinsic contributions to the damping and linewidth were discovered in the 1960's and 1970's, and are well described by theory, e.g. eddy-current damping[13,14], two-magnon scattering[15–17], the slow relaxer mechanism[18,19], or radiative damping[20,21], interest in these mechanisms has been re-ignited recently[22,23]. Further contributions, such as spin-pumping, both extrinsic[24,25] and intrinsic[24,26], have been discovered more recently and are subject to extensive research[27–31] for spintronics application. Therefore, in order to allow a quantitative comparison to theoretical calculations for intrinsic damping, both the measurement and sample geometry must be designed to allow both the determination and possible minimization of all additional contributions to the measured damping.

In this study, we demonstrate methods to determine significant extrinsic contributions to the damping, which includes a measurement of the effective spin mixing conductance for both the pure elements and select alloys. By precisely accounting for all of these extrinsic contributions, we determine the intrinsic damping parameters of the binary alloys $Ni_xCo_{1-x}$, $Ni_xFe_{1-x}$ and $Co_xFe_{1-x}$ and compare them to the calculations by Mankovsky, et al.,[1], Turek, et al., and Starikov, et al.[2]. Furthermore, we present the concentration-dependence of the inhomogeneous linewidth broadening, which for most alloys shows exceptionally small values, indicative of the high homogeneity of our samples.

## 2 Samples and method

We deposited $Ni_xCo_{1-x}$, $Ni_xFe_{1-x}$ and $Co_xFe_{1-x}$ alloys of varying composition (all compositions given in atomic percent) with a thickness of 10 nm on an oxidized (001) Si substrate with a Ta(3 nm)/Cu(3 nm) seed layer and a Cu(3 nm)/Ta(3 nm) cap layer. In order to investigate interface effects, we also deposited multiple thickness series at 10 nm, 7 nm, 4 nm, 3 nm, and 2 nm of both the pure elements and select alloys. Structural characterization was performed using X-ray diffraction (XRD). Field swept vector-network-analyzer ferromagnetic resonance spectroscopy (VNA-FMR) was used in the out-of-plane geometry to determine the total damping parameter $\alpha_{tot}$. Further details of the deposition conditions, XRD, FMR measurement and fitting of the complex susceptibility to the measured $S_{21}$ parameter are reported in Ref [66].

An example of susceptibility fits to the complex $S_{21}$ data is shown in Fig. 1 (a) and (b). All fits were constrained to a 3× linewidth $\Delta H$ field window around the resonance field in order to minimize the influence of measurement drifts on the error in the susceptibility fits. The total damping parameter $\alpha_{tot}$ and the inhomogeneous linewidth broadening $\Delta H_0$ are then determined from a fit to the linewidth $\Delta H$ vs. frequency $f$ plot[22], as shown in Fig. 1 (c).

$$\Delta H = \frac{4\pi \alpha_{tot} f}{\gamma \mu_0} + \Delta H_0, \qquad (1)$$

where $\gamma = g\mu_B/\hbar$ is the gyro-magnetic ratio, $\mu_0$ is the vacuum permeability, $\mu_B$ is the Bohr-magneton, $\hbar$ is the reduced Planck constant, and $g$ is the spectroscopic $g$-factor reported in Ref [66].

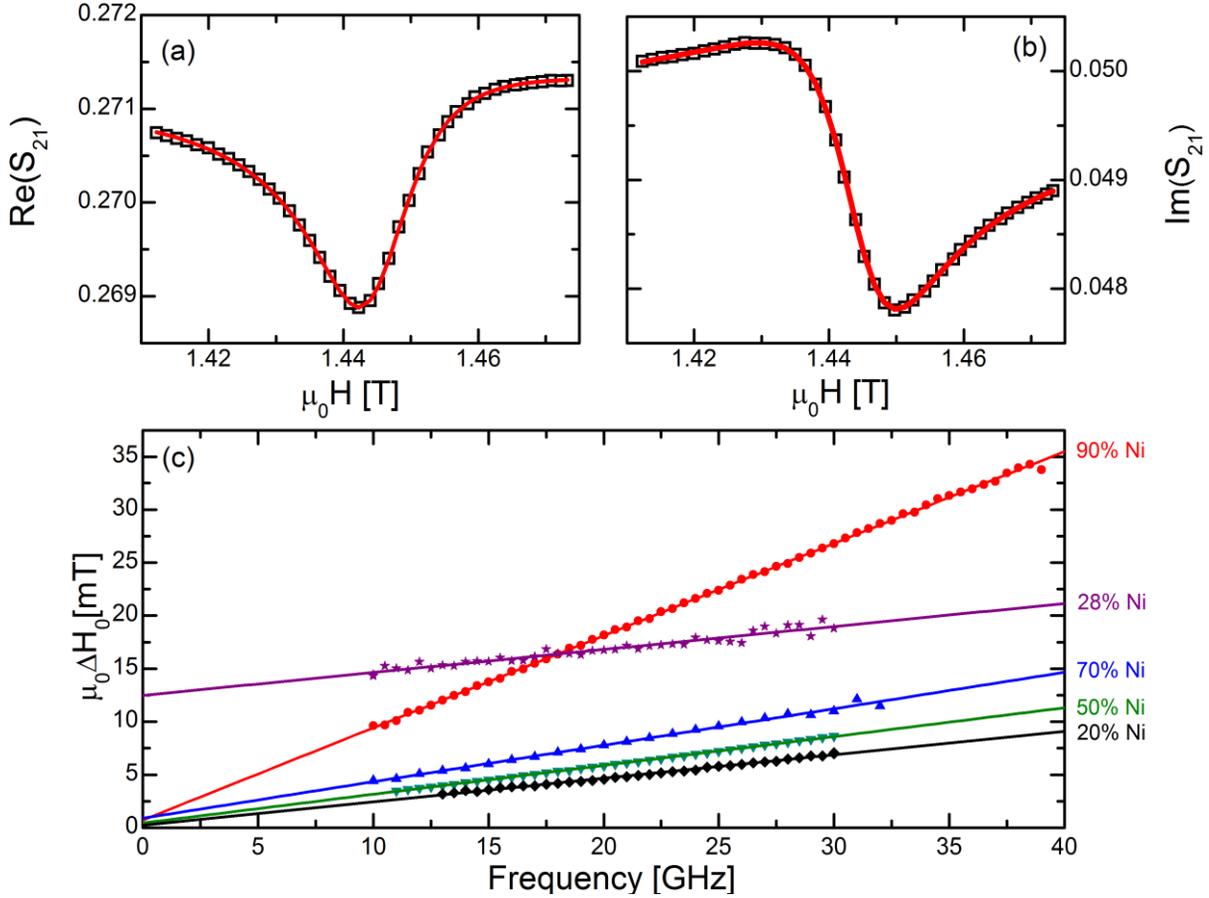

Figure 1: (a) and (b) show respectively the real and imaginary part of the $S_{21}$ transmission parameter (black squares) measured at 20 GHz with the complex susceptibility fit (red lines) for the $Ni_{90}Fe_{10}$ sample. (c) The linewidths from the susceptibility fits (symbols) and linear fits (solid lines) are plotted against frequency for different Ni-Fe compositions. Concentrations are denoted on the right-hand axis. The damping α and the inhomogeneous linewidth broadening $ΔH_0$ for each alloy can be extracted from the fits via Eq. (1).

## 3 Results

The first contribution to the linewidth we discuss is the inhomogeneous linewidth broadening $ΔH_0$, which is presumably indicative of sample inhomogeneity[32,33]. We plot $ΔH_0$ for all the alloy systems against the respective concentrations in Fig. 2. For all alloys, $ΔH_0$ is in the range of a few mT to 10 mT. There are only a limited number of reports for $ΔH_0$ in the literature with which to compare. For Permalloy ($Ni_{80}Fe_{20}$) we measure $ΔH_0 = 0.35$ mT, which is close to other reported values.[34] For the other $Ni_xFe_{1-x}$ alloys, $ΔH_0$ exhibits a significant peak near the fcc-to-bcc (face-centered-cubic to body-centered-cubic) phase transition at 30 % Ni, (see Fig. 2 (b)) which is easily seen in the raw data in Fig. 1 (c). We speculate that this increase of inhomogeneous broadening in the $Ni_xFe_{1-x}$ is caused by the coexistence of the bcc and fcc phases at the phase transition. However, the $Co_xFe_{1-x}$ alloys do not exhibit an increase in $ΔH_0$ at the equivalent phase transition at 70 % Co. This suggests that the bcc and fcc phases of $Ni_xFe_{1-x}$ tend to segregate near the phase transition, whereas the same phases for $Co_xFe_{1-x}$ remain intermixed throughout the transition.

One possible explanation for inhomogeneous broadening is magnetic anisotropy, as originally proposed in Ref. [35]. However, this explanation does not account for our measured dependence of $\Delta H_0$ on alloy concentration, since the perpendicular magnetic anisotropy, described in Ref [66] effectively exhibits opposite behavior with alloy concentration. For our alloys $\Delta H_0$ seems to roughly correlate to the inverse exchange constant[36,37], which could be a starting point for future investigation of a quantitative theory of inhomogeneous broadening.

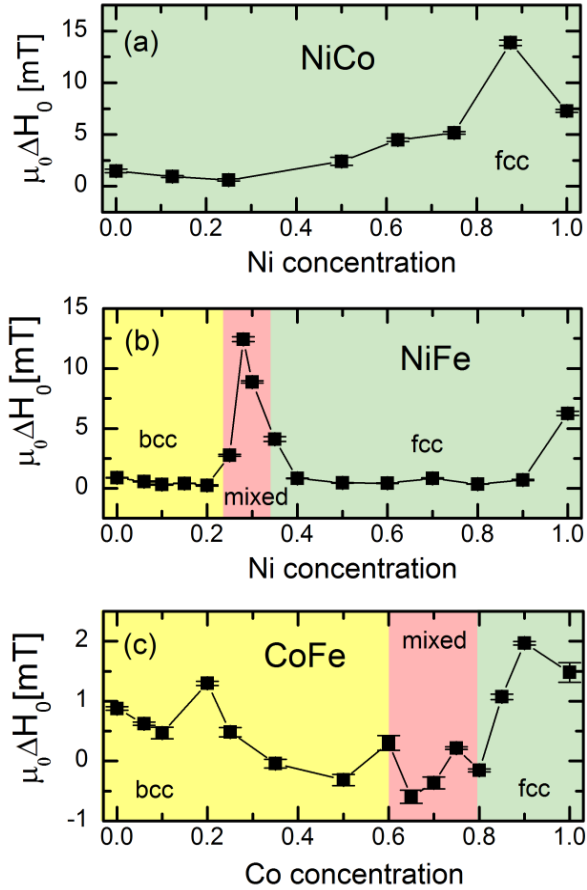

Figure 2: The inhomogeneous linewidth-broadening $\Delta H_0$ is plotted vs. alloy composition for (a) Ni-Co, (b) Ni-Fe and (c) Co-Fe. The alloy phases are denoted by color code described in Ref [66]

We plot the total measured damping $\alpha_{tot}$ vs. composition for $Ni_xCo_{1-x}$, $Ni_xFe_{1-x}$ and $Co_xFe_{1-x}$ in Fig. 3 (red crosses). The total damping of the $Ni_xCo_{1-x}$ system increases monotonically with increased Ni content. Such smooth behavior in the damping is not surprising owing to the absence of a phase transition for this alloy. In the $Ni_xFe_{1-x}$ system, $\alpha_{tot}$ changes very little from pure Fe to approximately 25 % Ni where the bcc to fcc phase transition occurs. At the phase transition, $\alpha_{tot}$ exhibits a step, increasing sharply by approximately 30 %. For higher Ni concentrations, $\alpha_{tot}$ increases monotonically with increasing Ni concentration. On the other hand, the $Co_xFe_{1-x}$ system shows a different behavior in the damping and displays a sharp minimum of $(2.3 \pm 0.1)\times10^{-3}$ at 25

% Co as previously reported[38]. As the system changes to an fcc phase (≈ 70 % Co), $\alpha_{tot}$ become almost constant.

We compare our data to previously published values in Table I. However, direct comparison of our data to previous reports is not trivial, owing to the variation in measurement conditions and sample characteristics for all the reported measurements. For example, the damping can depend on the temperature.[9,39] In addition, multiple intrinsic and intrinsic contributions to the total damping are not always accounted for in the literature. This can be seen in the fact that the reported damping in $Ni_{80}Fe_{20}$ (Permalloy) varies from α=0.0055 to α=0.04 at room temperature among studies. The large variation for these reported data is possibly the result of different uncontrolled contributions to the extrinsic damping that add to the *total* damping in the different experiments, e.g. spin-pumping[40–42], or roughness[41]. Therefore, the value for the intrinsic damping of $Ni_{20}Fe_{80}$ is expected to be at the low end of this scatter. Our measured value of α=0.0072 lies within the range of reported values. Similarly, many of our measured damping values for different alloy compositions lie within the range of reported values[22,43–48]. Our measured damping of the pure elements and the $Ni_{80}Fe_{20}$ and $Co_{90}Fe_{10}$ alloys is compared to room temperature values found in literature in Table 1, Columns 2 and 3. Column 5 contains theoretically calculated values.

Table 1: The total measured damping $\alpha_{tot}$ (Col. 2) and the intrinsic damping (Col. 4) for $Ni_{80}Fe_{20}$, $Co_{90}Fe_{10}$, and the pure elements are compared to both experimental (Col. 3) and theoretical (Col. 5) values from the literature. All values of the damping are at room temperature if not noted otherwise.

| Material | $\alpha_{tot}$ (this study) | Literature values | $\alpha_{int}$ (this study) | Calculated literature values |
|---|---|---|---|---|
| Ni | 0.029 (fcc) | 0.064[44] <br> 0.045[49] | 0.024 (fcc) | 0.017[9] (fcc) at 0K <br> 0.022[12] (fcc) at 0K <br> 0.013[1] (fcc) |
| Fe | 0.0036 (bcc) | 0.0019[44] <br> 0.0027[46] | 0.0025 (bcc) | 0.0013[9] (bcc) at 0K <br> 0.0010[12] (bcc) at 0K <br> 0.0012[1] (bcc) at 0K |
| Co | 0.0047 (fcc) | 0.011[44] | 0.0029 (fcc) | 0.0011[9] (hcp) at 0K <br> 0.00073[12] (hcp) at 0K <br> 0.001[1] (hcp) |
| $Ni_{80}Fe_{20}$ | 0.0073 (fcc) | 0.008[44] <br> 0.008-0.04[50] <br> 0.0078[48] <br> 0.007[51] <br> 0.006[52] <br> 0.006[47] <br> 0.0055[53] | 0.0050 (fcc) | 0.0046[2,54] (fcc) at 0K <br> 0.0039-0.0049[3] (fcc) at 0K |
| $Co_{90}Fe_{10}$ | 0.0048 (fcc) | 0.0043[44] <br> 0.0048[55] | 0.0030 (fcc) | |

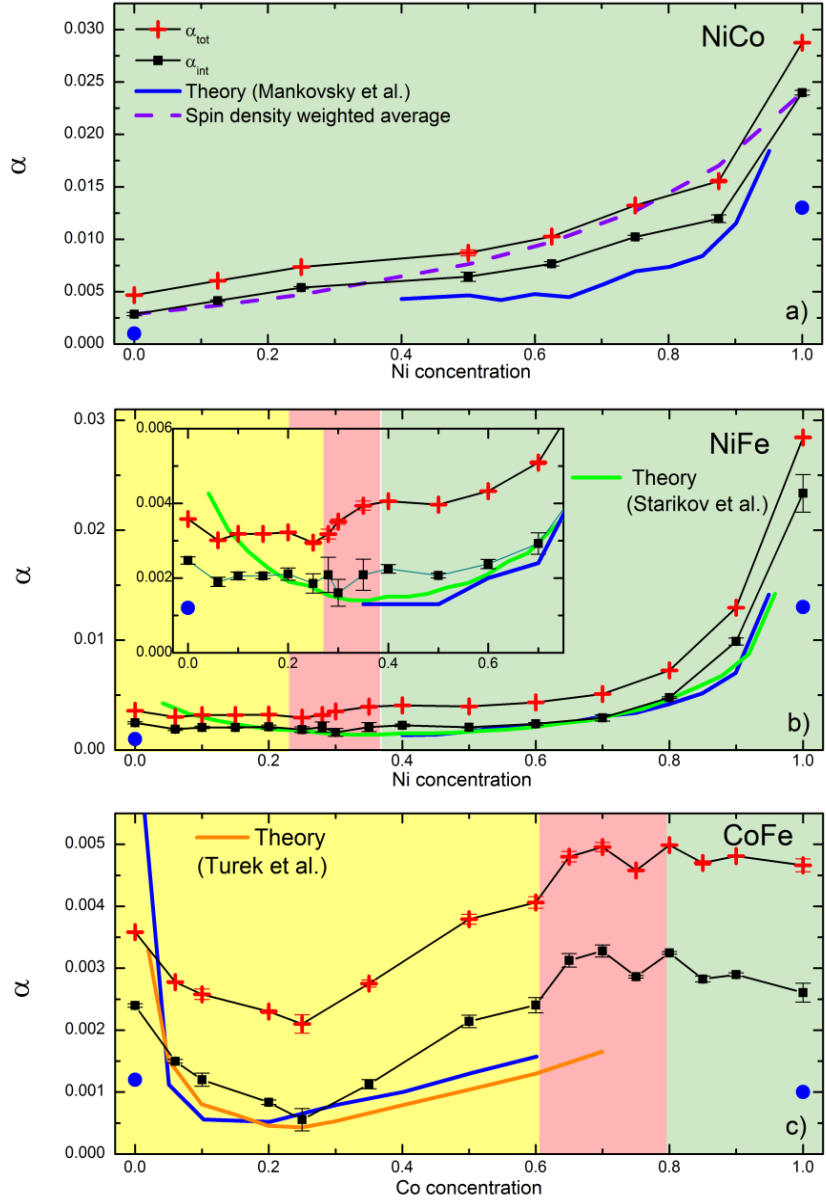

Figure 3: (color online) The measured damping $\alpha_{tot}$ of all the alloys is plotted against the alloy composition (red crosses) for (a) Ni-Co, (b) Ni-Fe and (c) Co-Fe (the data in (c) are taken from Ref.[38]). The black squares are the intrinsic damping $\alpha_{int}$ after correction for spin pumping and radiative contributions to the measured damping. The blue line is the intrinsic damping calculated from the Ebert-Mankovsky theory,[1] where the blue circles are the values for the pure elements at 300K. The green line is the calculated damping for the Ni-Fe alloys by Starikov, et al.[2] The inset in (b) depicts the damping in a smaller concentration window in order to better depict the small features in the damping around the phase transition. The damping for the Co-Fe alloys, calculated by Turek et al.[3] is plotted as the orange line. For the Ni-Co alloys the damping calculated by the spin density of the respective alloy weighted bulk damping[55] (purple dashed line).

This scatter in the experimental data reported in the literature and its divergence from calculated values of the damping shows the necessity to determine the intrinsic damping $\alpha_{int}$ by quantification of all extrinsic contributions to the measured total damping $\alpha_{tot}$.

The first extrinsic contribution to the damping that we consider is the radiative damping $\alpha_{rad}$, which is caused by inductive coupling between sample and waveguide, which results in energy flow from the sample back into the microwave circuit.[23] $\alpha_{rad}$ depends directly on the measurement method and geometry. The effect is easily understood, since the strength of the inductive coupling depends on the inductance of the FMR mode itself, which is in turn determined by the saturation magnetization, sample thickness, sample length, and waveguide width. Assuming a homogeneous excitation field, a uniform magnetization profile throughout the sample, and negligible spacing between the waveguide and sample, $\alpha_{rad}$ is well approximated by[23]

$$\alpha_{rad} = \frac{\gamma M_s \mu_0^2 \delta l}{16 Z_0 w_{cc}}, \qquad (2)$$

where $l$ (= 10 mm in our case) is the sample length on the waveguide, $w_{cc}$ (= 100μm) is the width of the co-planar wave guide center conductor and $Z_0$ (= 50 Ω), the impedance of the waveguide. Though inherently small for most thin films, $\alpha_{rad}$ can become significant for alloys with exceptionally small intrinsic damping and/or high saturation magnetization. For example, it plays a significant role (values of $\alpha_{rad} \approx 5 \times 10^{-4}$) for the whole composition range of the Co-Fe alloy system and the Fe-rich side of the Ni-Fe system. On the other hand, for pure Ni and Permalloy ($Ni_{80}Fe_{20}$) $\alpha_{rad}$ comprises only 3 % and 5 % of $\alpha_{tot}$, respectively.

The second non-negligible contribution to the damping that we consider is the interfacial contribution to the measured damping, such as spin-pumping into the adjacent Ta/Cu bilayers. Spin pumping is proportional to the reciprocal sample thickness as described in[24]

$$\alpha_{sp} = \frac{2 g_{eff}^{\uparrow\downarrow} \mu_B g}{4\pi M_s t}. \qquad (3)$$

The spectroscopic $g$-factor and the saturation magnetization $M_s$ of the alloys were reported in Ref [66] and the factor of 2 accounts for the presence of two nominally identical interfaces of the alloys in the cap and seed layers. In Fig. 4 (a)-(c) we plot the damping dependence on reciprocal thickness $1/t$ for select alloy concentrations, which allows us to determine the effective spin mixing conductance $g_{eff}^{\uparrow\downarrow}$ through fits to Eq. (3). The effective spin mixing conductance contains details of the spin transport in the adjacent non-magnetic layers, such as the interfacial spin mixing conductance, both the conductivity and spin diffusion for all the non-magnetic layers with a non-negligible spin accumulation, as well as the details of the spatial profile for the net spin accumulation.[56,57] The values of $g_{eff}^{\uparrow\downarrow}$, are plotted versus the alloy concentration in Fig. 4 (d), and are in the range of previously reported values for samples prepared under similar growth conditions[55–59]. Intermediate values of $g_{eff}^{\uparrow\downarrow}$ are determined by a guide to the eye interpolation [grey lines, Fig. 4 (d)] and $\alpha_{sp}$ is calculated for all alloy concentrations utilizing those interpolated values.

The data for $g_{eff}^{\uparrow\downarrow}$ in the $Ni_xFe_{1-x}$ alloys shows approximately a factor two increase of $g_{eff}^{\uparrow\downarrow}$ between Ni concentrations of 30 % Ni and 50 % Ni, which we speculate to occur at the fcc to bcc phase transition around 30 % Ni. According to this line of speculation, the previously mentioned step increase in the measured total damping at the $Ni_xFe_{1-x}$ phase transition can be fully attributed to the increase in spin pumping at the phase transition. In $Co_xFe_{1-x}$, the presence of a step in $g_{eff}^{\uparrow\downarrow}$ at the phase transition is not confirmed, given the measurement precision, although we do observe an increase in the effective spin mixing conductance when transitioning from the bcc to fcc phase. The concentration dependence of $g_{\text{eff}}^{\uparrow\downarrow}$ requires further thorough investigation and we therefore restrict ourselves to reporting the experimental findings.

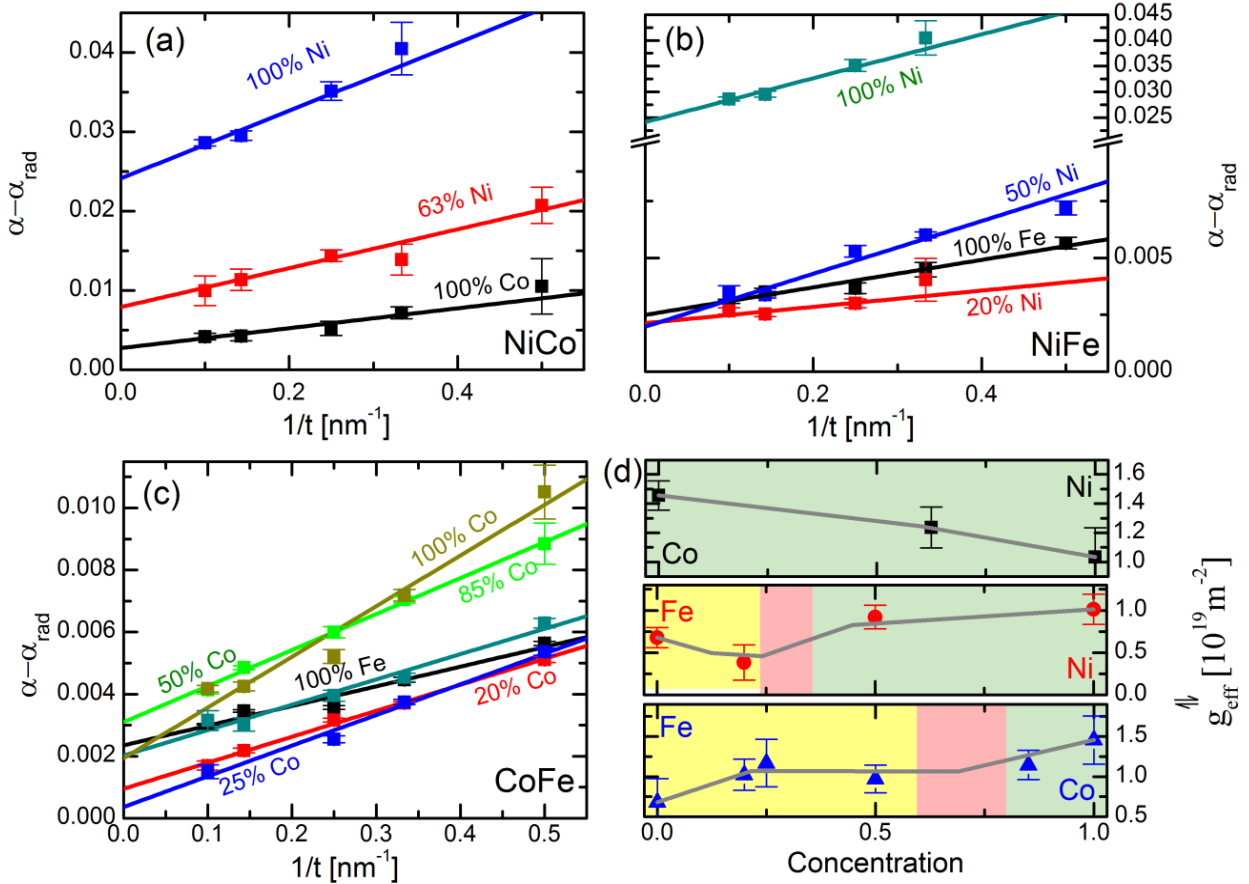

Figure 4: The damping for the thickness series at select alloy compositions vs. $1/t$ for (a) Ni-Co, (b) Ni-Fe and (c) Co-Fe (data points, concentrations denoted in the plots), with linear fits to Eq. (3) (solid lines). (d) The extracted effective spin mixing conductance $g_{\text{eff}}^{\uparrow\downarrow}$ for the measured alloy systems, where the gray lines show the linear interpolations for intermediate alloy concentrations. The data for the Co-Fe system are taken from Ref.[38].

Eddy-current damping[13,14] is estimated by use of the equations given in Ref. [23] for films 10 nm thick or less. Eddy currents are neglected because they are found to be less than 5 % of the total damping. Two-magnon scattering is disregarded because the mechanism is largely excluded in the out-of-plane measurement geometry[15–17]. The total measured damping is therefore well approximated as the sum

$$\alpha_{\text{tot}} \cong \alpha_{\text{int}} + \alpha_{\text{rad}} + \alpha_{\text{sp}}, \qquad (4)$$

We determine the intrinsic damping of the material by subtracting $\alpha_{\text{sp}}$ and $\alpha_{\text{rad}}$ from the measured total damping, as shown in Fig. 3.

The intrinsic damping increases monotonically with Ni concentration for the $Ni_xCo_{1-x}$ alloys. Indicative of the importance of extrinsic sources of damping, $\alpha_{int}$ is approximately 40 % smaller than $\alpha_{tot}$ for the Fe-rich alloy, though the difference decreases to only 15 % for pure Ni. This behavior is expected, given that both $\alpha_{rad}$ and $\alpha_{sp}$ are proportional to $M_s$. A comparison of $\alpha_{int}$ to the calculations by Mankovsky, et al.,[1] shows excellent quantitative agreement to within 30 %. Furthermore, we compare $\alpha_{int}$ of the $Ni_xCo_{1-x}$ alloys to the spin density weighted average of the intrinsic damping of Ni and Co [purple dashed line in Fig. 3 (a)] , which gives good agreement with our data, as previously reported.[55]

$\alpha_{int}$ for $Ni_xFe_{1-x}$ (Fig. 3 (b)) also increases with Ni concentration after a small initial decrease from pure Fe to the first $Ni_xFe_{1-x}$ alloys. The step increase found in $\alpha_{tot}$ at the bcc to fcc phase transition is fully attributed to $\alpha_{sp}$, as detailed in the previous section, and therefore does not occur in $\alpha_{int}$. Similar to the $Ni_xCo_{1-x}$ system $\alpha_{int}$ is significantly lower than $\alpha_{tot}$ for Fe-rich alloys. With in error bars, a comparison to the calculations by Mankovsky, et al.[1] (blue line) and Starikov, et al.[2] (green line) exhibit excellent agreement in the fcc phase, with marginally larger deviations in the Ni rich regime. Starikov, et al.[2] calculated the damping over the full range of compositions, under the assumption of continuous fcc phase. This calculation deviates further from our measured $\alpha_{int}$ in the bcc phase exhibiting qualitatively different behavior.

As previously reported, the dependence of $\alpha_{int}$ on alloy composition in the $Co_xFe_{1-x}$ alloys exhibits strongly non-monotonic behavior, differing from the two previously discussed alloys.[38] $\alpha_{int}$ displays a minimum at 25 % Co concentration with a, for conducting ferromagnets unprecedented, low value of $\alpha_{int} = (5 \pm 1.8) \times 10^{-4}$. With increasing Co concentration, $\alpha_{int}$ grows up to the phase transition, at which point it increases by 10 % to 20 % until it reaches the value for pure Co. It was shown that $\alpha_{int}$ scales with the density of states (DOS) at the Fermi energy $n(E_F)$ in the bcc phase[38], and the DOS also exhibits a sharp minimum for $Co_{25}Fe_{75}$. This scaling is expected[60,61] if the damping is dominated by the breathing Fermi surface process. With the breathing surface model, the intraband scattering that leads to damping directly scales with $n(E_F)$. This scaling is particularly pronounced in the Co-Fe alloy system due to the small concentration dependence of the spin-orbit coupling on alloy composition. The special properties of the $Co_xFe_{1-x}$ alloy system are discussed in greater detail in Ref.[[38]].

Comparing $\alpha_{int}$ to the calculations by Mankovsky et al.[1], we find good quantitative agreement with the value of the minimum. However, the concentration of the minimum is calculated to occur at approximately 10 % to 20% Co, a slightly lower value than 25 % Co that we find in this study. Furthermore, the strong concentration dependence around the minimum is not reflected in the calculations. More recent calculations by Turek et al.[3], for the bcc $Co_xFe_{1-x}$ alloys [orange line in Fig. 3 (c)] find the a minimum of the damping of $4 \times 10^{-4}$ at 25 % Co concentration in good agreement with our experiment, but there is some deviation in concentration dependence of the damping around the minimum. Turek et al.[3] also reported on the damping in the $Ni_xFe_{1-x}$ alloy system, with similar qualitative and quantitative results as the other two presented quantitative theories[1,2] and the results are therefore not plotted in Fig. 3 (b) for the sake of comprehensibility of the figure. For both $Ni_xFe_{1-x}$ and the $Co_xFe_{1-x}$ alloys, the calculated spin density weighted intrinsic damping of the pure elements (not plotted) deviates significantly from the determined intrinsic damping of the alloys, in contrary to the good agreement archived for the $Co_xNi_{1-x}$ alloys. We speculate that this difference between the alloy systems is caused by the non-monotonous dependence of the density of states at the Fermi Energy in the $Co_xFe_{1-x}$ and $Ni_xFe_{1-x}$ systems.

Other calculated damping values for the pure elements and the $Ni_{80}Fe_{20}$ and $Co_{90}Fe_{10}$ alloys are compared to the determined intrinsic damping in Table 1. Generally, the calculations underestimate the damping significantly, but our data are in good agreement with more recent calculations for Permalloy ($Ni_{80}Fe_{20}$).

It is important to point out that none of the theories considered here include thermal fluctuations. Regardless, we find exceptional agreement with the calculations to $\alpha_{int}$ at intermediate alloy concentrations. We speculate that the modeling of atomic disorder in the alloys in the calculations, by the coherent potential approximation (CPA) could be responsible for this exceptional agreement. The effect of disorder on the electronic band structure possibly dominates any effects due to nonzero temperature. Indeed, both effects cause a broadening of the bands due to enhanced momentum scattering rates. This directly correlates to a change of the damping parameter according to the theory of Gilmore and Stiles[9]. Therefore, the inclusion of the inherent disorder of solid-solution alloys in the calculations by Mankovsky et al[1] mimics the effects of temperature on damping to some extent. This argument is corroborated by the fact that the calculations by Mankovsky et al[1] diverge for diluted alloys and pure elements (as shown in Fig. 2 (c) for pure Fe), where no or to little disorder is introduced to account for temperature effects. Mankovsky et al.[1] performed temperature dependent calculations of the damping for pure bcc Fe, fcc Ni and hcp Co and the values for 300 K are shown in Table 1 and Fig. 3. These calculations for $\alpha_{int}$ at a temperature of 300 K are approximately a factor of two less than our measured values, but the agreement is significantly improved relative to those obtained by calculations that neglect thermal fluctuations.

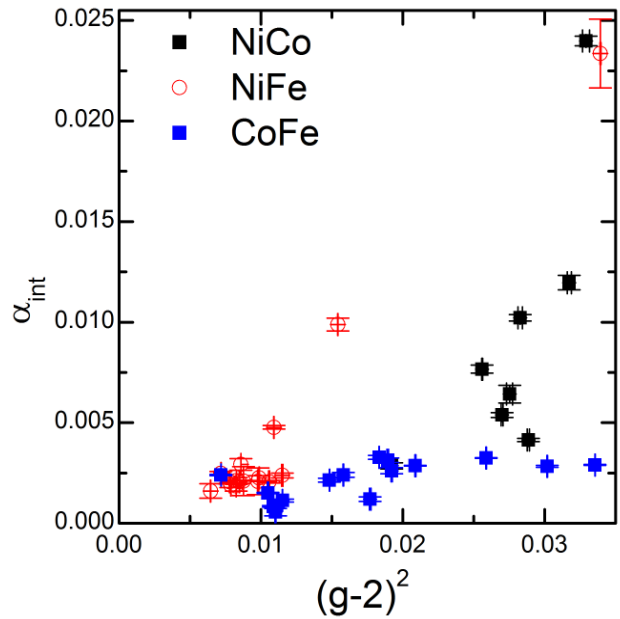

Figure 5: The intrinsic damping $\alpha_{int}$ is plotted against $(g-2)^2$ for all alloys. We do not observe a proportionality between $\alpha_{int}$ and $(g-2)^2$.

Finally, it has been reported[45,64] that there is a general proportionality between $\alpha_{int}$ and $(g-2)^2$, as contained in the original microscopic BFS model proposed by Kambersky.[62] To examine this relationship, we plot $\alpha_{int}$ versus $(g-2)^2$ (determined in Ref [66]) for all samples measured here in Figure 5. While some samples with large values for $(g-2)^2$ also exhibit large $\alpha_{int}$, this is not a general trend for all the measured samples. Given that the damping is not purely a function of the spin-orbit strength, but also depends on the details of the band structure, the result in Fig. 5 is expected. For example, the amount of band overlap will determine the amount of interband transition leading to that damping channel. Furthermore, the density of states at the Fermi energy will affect the intraband contribution to the damping[9,10]. Finally, the ratio of inter- to intra-band scattering that mediates damping contributions at a fixed temperature (RT for our measurements) changes for different elements[9,10] and therefore with alloy concentration. None of these factors are necessarily proportional to the spin-orbit coupling. Therefore, we conclude that this simple relation, which originally traces to an order of magnitude estimate for the case of spin relaxation in semiconductors[65], does not hold for all magnetic systems in general.

## 4 Summary

We determined the damping for the full composition range of the binary 3$d$ transition metal alloys Ni-Co, Ni-Fe, and Co-Fe and showed that the measured damping can be explained by three contributions to the damping: Intrinsic damping, radiative damping and damping due to spin pumping. By quantifying all extrinsic contributions to the measured damping, we determine the intrinsic damping over the whole range of alloy compositions. These values are compared to multiple theoretical calculations and yield excellent qualitative and good quantitative agreement for intermediate alloy concentrations. For pure elements or diluted alloys, the effect of temperature seems to play a larger role for the damping and calculations including temperature effects give significantly better agreement to our data. Furthermore, we demonstrated a compositional dependence of the spin mixing conductance, which can vary by a factor of two. Finally, we showed that the often postulated dependence of the damping on the $g$-factor does not apply to the investigated binary alloy systems, as their damping cannot be described solely by the strength of the spin-orbit interaction.